\documentclass[conference]{IEEEtran}
\IEEEoverridecommandlockouts
\usepackage{cite}
\usepackage{amsmath,amssymb,amsfonts}
\usepackage{algorithmic}
\usepackage{graphicx}
\usepackage{textcomp}
\usepackage{xcolor}
\usepackage{color}
\usepackage{comment}
\usepackage{url}
\def\BibTeX{{\rm B\kern-.05em{\sc i\kern-.025em b}\kern-.08em
    T\kern-.1667em\lower.7ex\hbox{E}\kern-.125emX}}

\newtheorem{definition}{Definition}
    
\begin{document}

\title{Population Graph Cross-Network Node Classification for Autism Detection Across Sample Groups
}


\author{\IEEEauthorblockN{Anna Stephens, Francisco Santos, Pang-Ning Tan, and Abdol-Hossein Esfahanian}
\IEEEauthorblockA{\textit{Dept. of Computer Science and Engineering} \\
\textit{Michigan State University}\\
East Lansing, United States \\
\{steph496, santosf3, ptan, esfahanian\}@msu.edu}
}

\maketitle

\begin{abstract}

Graph neural networks (GNN) are a powerful tool for combining imaging and non-imaging medical information for node classification tasks. Cross-network node classification extends GNN techniques to account for domain drift, allowing for node classification on an unlabeled target network. In this paper we present OTGCN, a powerful, novel approach to cross-network node classification. This approach leans on concepts from graph convolutional networks to harness insights from graph data structures while simultaneously applying strategies rooted in optimal transport to correct for the domain drift that can occur between samples from different data collection sites. This blended approach provides a practical solution for scenarios with many distinct forms of data collected across different locations and equipment. We demonstrate the effectiveness of this approach at classifying Autism Spectrum Disorder subjects using a blend of imaging and non-imaging data.
\end{abstract}

\begin{IEEEkeywords}
autism, graph, transfer learning, cross-network node classification 
\end{IEEEkeywords}

\section{Introduction}


Autism Spectrum Disorder (ASD) refers to a condition characterised by specific communication impairments, restricted interests, and repetitive behaviours \cite{lord2020autism}. As the disorder typically presents itself early in life, early and accurate detection can help reduce the severity of many lifelong symptoms. Towards this end, automated techniques based on machine learning and deep learning have been developed for the early detection of ASD. Specifically, these techniques have been applied to a wide range of subject data including detailed subject screening data \cite{erkan2019autism}, videos of subject movements \cite{zunino2018video}, and Magnetic Resonance Imaging (MRI) data \cite{li2021braingnn}. Given the diverse modalities of data available, it is likely that the best detection results can be  achieved by combining the data from different sources.  

Previous works have established an effective method to combine various modalities of data using graph structures because of their flexible yet powerful representation \cite{tong2017multi} \cite{zeng2021multi}. These approaches make use of population graphs, where the nodes represent individuals and edges are generally defined with some similarity measure. In several prior ASD research, the node features are generated from image information while the edges consider a combination of subject information (sex, age, etc) in addition to image similarity \cite{parisot2017spectral} \cite{parisot2018disease} \cite{kazi2019inceptiongcn} \cite{jiang2020higcn}. Once these graphs are constructed, detecting ASD becomes a \emph{node classification} problem.

\begin{figure}[t]
\centerline{\includegraphics[scale=0.48]{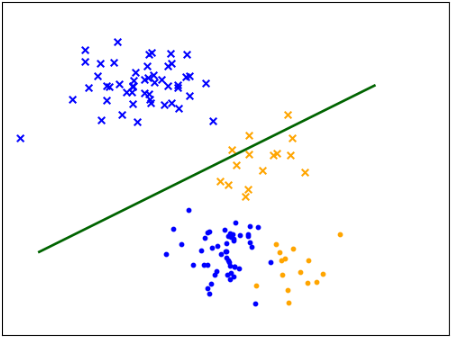}}
\caption{An example illustrating the perils of failing to account for concept drift in multi-source data. Assume a binary node classification task where the two classes are represented  as x's and o's, respectively. The blue points denote the dataset for the source domain while the orange points denote the dataset for the target domain. Observe that a decision boundary constructed from the blue dataset will have reduced accuracy when applied to the orange dataset due to concept drift between the source and target networks. }
\label{fig_gcn_embed}
\end{figure}

Similar to other diagnosis problems, one of the major challenges in applying machine learning to ASD detection is the limited amount of labeled data available. This has led to growing interest in utilizing labeled data from multiple sites to train the machine learning model. One limitation of these approaches is that they fail to consider a more realistic scenario in which there may be little or no labeled data associated with the population of interest. For example, a model may be trained on a research dataset but needs to be applied to another location with a different imaging equipment or patient demographics that were not well represented in the training data. This problem can be addressed using an approach generally known as domain adaptation or transfer learning \cite{tan2018survey}. Within the context of node classification tasks, it is also referred to as \emph{cross-network node classification} (CNNC) approach. CNNC assumes the availability of a sufficiently large number of labeled nodes in a source network and an unlabeled target network, whose node labels are to be predicted accurately. 

This paper focuses on addressing the CNNC task for ASD detection. We use the popular ABIDE \cite{dataset_abide} dataset, which contains both resting-state functional Magnetic Resonance Imaging (fMRI) information and phenotypic data such as age, sex, and screening results. Samples in the ABIDE dataset were collected from several different collection sites, which were then divided into 2 groups to form the source and target networks for our experimental studies.

CNNC has two major challenges when applied to the ABIDE dataset. The first is extracting relevant information from the networks for ASD detection. In this work we will learn a graph embedding of the source and target networks via a graph convolutional network (GCN) \cite{gcn_kipf_welling}. A GCN layer is capable of learning a node embedding which contains information about a node and its immediate neighbors. This method relies on the homophily principle \cite{mcpherson2001birds}, which states that similar individuals tend to form neighborhoods within a graph. Therefore, in a graph with high homophily we can improve a node embedding by adding information from it's neighborhood. 

The second major challenge is to handle potential concept drift between the source and target networks. As the separate data collection sites may have different fMRI imaging equipment and procedural differences, this may introduce some discrepancies or "drift" between the two networks. The presence of concept drift often leads to poor results if a classifier is simply trained on the source dataset and applied as it is to the target dataset. Fig. \ref{fig_gcn_embed} demonstrates the challenge of CNNC when the learned node embedding does not account for such distributional shift. A decision boundary learned from the source dataset is likely to incorrectly classify a significant portion of the target dataset. 

In this paper we introduce OTGCN, a novel approach to address the CNNC task for ASD detection.  Our proposed approach combines graph neural network with optimal transport (OT) to handle the drift between the source and target networks. OT is a method for mapping a transportation between two distributions. We will use OT to map the learned source representation to the target representation. This strategy allows us to train an accurate model for classifying the target nodes. Experimental results on the ABIDE dataset demonstrate the effectiveness of our approach at diagnosing ASD across different collection sites compared to state of the art CNNC methods.

\section{Related Works}

\subsection{Machine Learning Approaches to ASD Detection}

ASD detection is a task that lends itself to a wide variety of approaches. Zunino et al \cite{zunino2018video} employed computer vision approaches for automatic early detection of ASD by evaluating videos of subject movements when grasping a bottle.  They then applied recurrent neural networks to distinguish subjects who have ASD from those who do not. Erkan and Thanh \cite{erkan2019autism} analyzed detailed screening data collected from mobile app surveys using traditional machine learning methods such as k-nearest neighbor, support vector machines, and random forests to diagnose subjects with ASD. Yuan et al. \cite{yuan2016autism} applied natural language processing (NLP) techniques to analyze hand written medical forms of potential ASD patients while Carette et al. \cite{carette2018automatic} performed ASD detection on eye tracking data using long-short term memory (LSTM) networks. 

The majority of the recent works in this area, however, focuses on using MRI data \cite{li2021braingnn} \cite{nogay2020machine} along with other subject information such as sex, age, and screening results \cite{parisot2017spectral} \cite{parisot2018disease} \cite{jiang2020higcn} \cite{kazi2019inceptiongcn} for screening potential ASD patients. For example, Li et al. \cite{li2021braingnn} presented a graph neural network approach to find biomarkers that can be used to detect ASD while Parisot et al. \cite{parisot2017spectral} employed a graph convolutional network (GCN) to perform the detection using both fMRI imaging and non-imaging phenotypic data. Similar to these works, our paper focuses on using a combination of fMRI and other subject data, though our approach is also applicable to blend other forms of data given the representation power of graph neural networks. 

\subsection{Machine Learning on Multi-site fMRI Data}

Previous works on diagnosing brain-related problems using fMRI data from multiple sites generally fall into two major categories---transfer learning and federated learning. 

Transfer learning \cite{tan2018survey} is a machine learning approach that enables prediction models trained from a given data domain (known as the \emph{source} domain) to be applied to another domain (known as the \emph{target} domain). Such a domain adaptation approach can be used even if the target domain has no labeled training data. 
Previous works on applying transfer learning to fMRI data can be found in
\cite{mrida_cheng2017multi} \cite{mrida_wang2020multi} \cite{mrida_guan2021multi}. However, these approaches are mostly designed for using only the image information and do not consider more complex data structures or the use of additional non-imaging information.


In contrast, federated learning \cite{zhang2021survey} 
is designed for training prediction models in a collaborative fashion on decentralized datasets. The approach assumes restricted or indirect access to the source dataset and direct access to a partially labeled target dataset. There has been a few works applying federated learning approaches to multi-site fMRI data \cite{li2020multi} \cite{wang2022metateacher}, but none of them consider non-imaging information.

\subsection{Cross-Network Node Classification (CNNC)}
There are a handful of studies focusing on applying domain adaptation approaches to multi-network data using graph neural networks. For node classification, these approaches are also collectively known as cross-network node classification (CNNC). For example, Shen et al. \cite{cnnc_cdne} presents an approach called CDNE that uses maximum mean discrepancy (MMD) loss to learn graph embeddings for the source and target networks before sending those embedding to a shared classifier. The authors subsequently extended the approach to ACDNE in \cite{cnnc_acdne} by adding an adversarial domain adaptation component to the framework.
AdaGCN \cite{cnnc_adagcn} combines conventional graph convolutional network (GCN) layers with adversarial domain adaptation, while UdaGCN \cite{cnnc_udagcn} uses multiple GCNs with both  graph adjacency and positive point-wise mutual information (PPMI) matrices to learn an improved combined feature embedding of the nodes. A more recent work, ASN \cite{cnnc_asn} combines the node embedding from UdaGCN with the adversarial domain adaptation approach from AdaGCN to address the CNNC problem. While all of these approaches were designed for CNNC tasks, they have not been tested on the medical dataset nor with a combination involving imaging and non-imaging data. 

\section{Preliminaries}

\subsection{{Problem Statement}}

Let $\mathcal{G}(V,E,X,Y)$ be an attributed graph, with a node set $V$, edge set $E \subseteq V \times V$, node feature matrix $X$, and node label $Y$. Let $A$ denote the adjacency matrix representation of $E$, where $A_{ij} > 0$ if $(v_i,v_j) \in E$ and 0 otherwise. In a domain adaptation setting, we assume there exists a source graph, $\mathcal{G}_s(V_s,E_s,X_s,Y_s)$, where the node labels in $Y_s$ are known, and a target graph, $\mathcal{G}_t(V_t,E_t,X_t,Y_t)$, where the node labels in $Y_t$ are unknown. The adjacency matrices corresponding to the source and target graphs are denoted as $A_s \in \mathbb{R}^{n_s \times n_s}$ and $A_t \in \mathbb{R}^{n_t \times n_t}$, respectively,  
where $n_s$ and $n_t$ are the corresponding number of source and target nodes. 
 
We further assume that both graphs have identical features, i.e., $X_s \in \mathbb{R}^{n_s\times m}$ and $X_t \in \mathbb{R}^{n_t\times m}$. Both graphs are also assumed to have the same set of class labels, i.e., $Y_s \in \{0,1,\cdots,k-1\}^{n_s}$ and $ Y_t \in \{0,1,\cdots,k-1\}^{n_t}$, where $k$ is the number of classes. For ASD detection, we consider a binary node classification problem with $k = 2$. At times, we also consider the combined graph $\mathcal{G}_c(V_c,E_c,X_c,Y_c)$, where $V_c = V_s \cup V_t$, $E_c = E_s \cup E_t$, $X_c = [X_s; X_t] \in \mathbb{R}^{n \times m}$, $A_c = [A_s \ \mathbf{0}; \ \mathbf{0}^T \ A_t] \in \mathbb{R}^{n \times n}$, $Y_c \in \{0,1,\cdots,k-1\}^{n}$, $n= n_s + n_t$, and $\mathbf{0}$ is an $n_s \times n_t$ null matrix. 

\begin{definition}[Cross Network Node Classification (CNNC)]
Given a source graph, $\mathcal{G}_s = (V_s, E_s, X_s,Y_s)$ and target graph, $\mathcal{G}_t = (V_t,E_t,X_t,Y_t)$, our goal is to learn a target function, $f: V \rightarrow \{0,1,\cdots,k-1\}$, that accurately classifies the labeled nodes in $Y_s$ as well as the unlabeled nodes in $Y_t$.    
\end{definition}


\subsection{Graph Convolutional Network (GCN)}
Graph convolutional networks \cite{gcn_kipf_welling} employ a message passing strategy to succinctly capture both 
the node features and graph structure information when  learning the feature embedding of a node in the graph. Specifically, each ``message" corresponds to the feature embedding information of a node, which will be passed to all of its immediate neighbors. By aggregating the features gathered from the neighbors, a new embedding for the node will be generated.

The message passing strategy can be formally stated as follows. Given an adjacency matrix $A$ and node feature matrix $X$, the feature embedding of the nodes in layer $l+1$, $H^{(l+1)}$, can be computed based on its feature embedding at its previous layer, $H^{(l)}$, as follows: 
\begin{equation}
H^{(l+1)} = \sigma(\tilde{D}^{-\frac{1}{2}} \tilde{A} \tilde{D}^{-\frac{1}{2}} H^{(l)} W^{(l)})
\label{eqn:gcn}
\end{equation}
$$\tilde{A} = A + I$$
$$\tilde{D}_{ii} = \sum_j \tilde{A}_{ij} $$
where $H^{(0)}=X$ (i.e., original node features), $\sigma(\cdot)$ is the ReLU activation function, and $\tilde{D}$ is a diagonal matrix containing the sum of the edge weights associated with each node in the graph.

\subsection{Optimal Transport}

\begin{figure}[t]
\centerline{\includegraphics[scale=0.4]{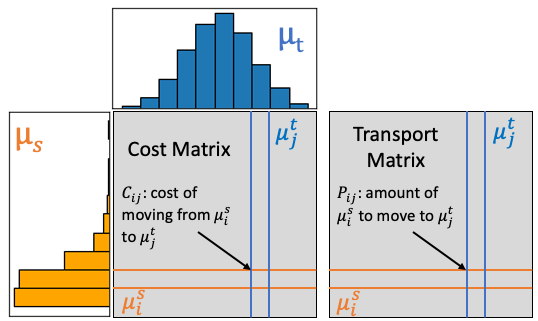}}
\caption{Illustration of the cost and transport plan matrices of optimal transport}
\label{fig_ot}
\end{figure}
\begin{figure*}[!th]
\centerline{\includegraphics[width=\textwidth]{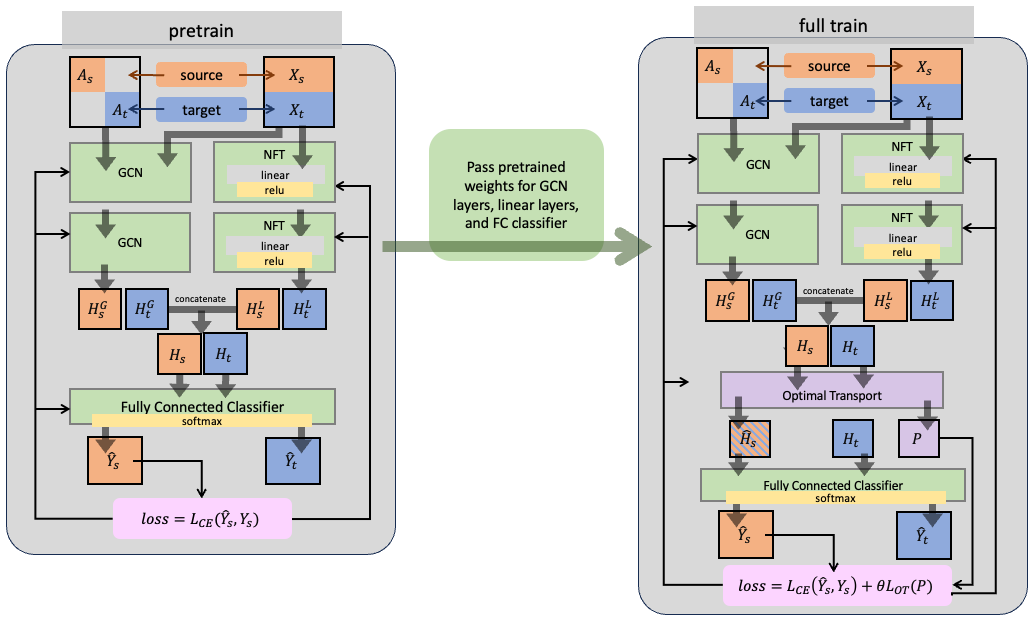}}
\caption{High level architecture of our model. Combined source and target datasets are fed to two GCN layers and two NFT layers to learn two distinct embeddings. In pretraining (left) these embeddings are concatenated and sent directly to a fully connected classifier. Model weights are then updated using just cross entropy loss. In OT training (right) concatenated embeddings are routed to an OT layer before being sent to the classifier. The OT layer replaces the source embedding with a transported version and then sends the target embedding and the transported source embedding to the classifier. At this point the model weights are updated with a combination of cross entropy and OT losses.  }
\label{fig_arc}
\end{figure*}
Optimal Transport (OT) provides a principled approach for comparing two probability distributions by finding the least costly way to reshape one of the distributions into the other while incorporating their geometric information. The original OT problem was proposed by Monge \cite{monge1781memoire}, with its modern formulation being attributed to Kantorovich \cite{kantorovich2006translocation}.
Given two marginal distributions $\mu_s \in \mathbb{R}^{n_s}$ and $\mu_t \in \mathbb{R}^{n_t}$ let $C\in \mathbb{R}^{n_s \times n_t}$ be a cost matrix, where $C_{ij}$ is the cost of transporting one unit from $\mu^s_i$ to $\mu^t_j$. Consider a transportation plan matrix $P \in \mathbb{R}^{n_s \times n_t} $, where $p_{ij}$ is the proportion of probability mass $\mu^s_i$ that should be moved to $\mu^t_j$. 

The optimal transport (OT) problem seeks to find a transportation plan $P$ that minimizes the total transportation cost. The minimum transportation cost $W(\mu_s,\mu_t)$ is also called the Wasserstein distance or the earth-mover distance. 

\begin{equation}
    W(\mu_s,\mu_t) := \min_{P \in U(\mu_s,\mu_t)} \langle P, C \rangle_F
    \label{eq:discrete_OT}
\end{equation}
$$
    U(\mu_s,\mu_t):=\{P\in \mathbb{R}_+^{n_s \times n_t} | P1_{n_t}=\mu_s, P^T1_{n_s}=\mu_t\}
$$

The discrete OT formulation \cite{ot_kantorovitch} shown above can be solved as a linear programming problem. However, it is computationally expensive $\mathcal{O}((n+m)nm\log(n+m))$, unstable, and is not guaranteed to have unique solution. Fortunately these problems can be addressed by employing the Sinkhorn distance $W_\lambda$ \cite{sinkhorn_cuturi} shown below, which utilizes an entropy regularization function $H(P)$ to 
accelerate the OT computation. Specifically,
$W_\lambda(\mu_s,\mu_t)$ and $P^\lambda$ can be solved using the well-known Sinkhorn algorithm as established by Cuturi \cite{sinkhorn_cuturi}.
\begin{eqnarray}
W_\lambda(\mu_s,\mu_t) &=& \langle P^\lambda , C \rangle \label{eq:OT_ent_reg} \\
P^\lambda &=& \underset{P \in U(\mu_s,\mu_t)}{\mathrm{argmin}} \langle P , C \rangle - \lambda H(P) \nonumber \\
U(\mu_s,\mu_t)&:=&\left\{P\in \mathbb{R}_+^{n_s \times n_t} \big| P1_{n_t}=\mu_s, P^T1_{n_s}=\mu_t\right\} \nonumber
\end{eqnarray}
where $H(P) = -\sum_{ij} P_{ij} \log P_{ij}$ is the entropy regularization and $\lambda$ is a user-specified regularization hyperparameter.

\section{Methodology}

We employed a combination of graph convolutional networks (GCN) with node feature transformation layers to learn the feature embedding of the nodes in the source and target graphs. We then performed optimal transport on the learned embedding of the source nodes to match the distribution of the learned embedding of the target nodes. Figure \ref{fig_arc} provides a high-level illustration of our proposed deep neural network architecture along with its training procedure. Details of the proposed architecture are described in the subsections below.

\subsection{Initial Node Embedding Construction \& Pretraining}

As previously noted, conventional GCN employs a message passing strategy to transmit information about the feature embedding of a node to its immediate neighbors. This allows each node to aggregate the feature information of its neighbors when constructing its own latent embedding. Unfortunately, in graphs with low homophily, it is possible that the neighborhood information obtained via message passing is of less value than the original feature information of the node itself, which was the case with the population graph constructed from the ABIDE dataset. However, due to its current formulation (see Eqn. \ref{eqn:gcn}), conventional GCN may not be able to attenuate the influence of the graph structure in comparison to the influence of the original features.

To overcome this challenge, we propose a modification to the graph convolutional network architecture to independently learn a nonlinear embedding of the original node features. We call this the \emph{node feature transformation} (NFT) layer in Fig \ref{fig_arc}. The NFT layer consists of a combination of linear layer plus ReLU activation functions to transform the original node features into their corresponding nonlinear embedding. The transformed features are then concatenated with the structural embedding of the nodes obtained from GCN for subsequent node classification. This strategy increases the flexibility of the model to capture the relative influence of the node features and homophily (i.e., neighborhood features) on the classification task.
As shown in the architecture diagram, the GCN layers are trained on combined adjacency and feature matrices of the source and target graphs, while the NFT layers are trained on the combined feature matrices.  The weights of the GCN and NFT layers are jointly updated during backpropagation.

The pretraining process of the deep neural network can be seen on the left side of Fig \ref{fig_arc}. The purpose of pretraining is to ensure that the full network with optimal transport (right side of Fig \ref{fig_arc}) can be seeded with a good set of initial weights.  During pretraining, we train the NFT and GCN layers to each produce their own feature embeddings, $H^L$ (output of NFT layer) and $H^G$ (output of GCN layer). The two embeddings are then concatenated and sent to a fully connected network for node classification. The network is trained to minimize the following cross-entropy loss function:
\begin{equation}
\textrm{Pretraining:} \ \ \mathcal{L}_{CE} = - \frac{1}{n_s} \sum_{x_i \in X_s} \sum^l_{j=0} Y_{ij} \log (\hat{Y}_{ij})
\label{eqn:crossent}
\end{equation}


\subsection{Domain Adaptation via Optimal Transport GCN}

We address the domain adaptation problem for cross network node classification (CNNC) by combining the pretrained network with an optimal transport layer as shown on the right side of Fig. \ref{fig_arc}. The optimal transport layer utilizes the Sinkhorn algorithm \cite{sinkhorn_cuturi} to learn the relevant transportation plan matrix $P$ that will transform the learned embedding of the source nodes to match the distribution of the target node embedding. 

The OT layer takes the source and target node embeddings along with the entropy regularization hyperparameter $\lambda$ as inputs and returns a new embedding of the source nodes, $\hat{H}_s$, which matches the distribution of the target nodes, $H_t$. This can be accomplished by solving the OT problem given in Eqn. \eqref{eq:OT_ent_reg} to learn the transportation plan $P$ and using the barycentric mapping approach in \cite{ot_map__courty} \cite{ot_map__ferradans} to transform the source node features. Specifically, for each source node $i$, its latent features will be transported to a new embedding as follows: 
$$\hat{H}_{s,i} = \underset{H\in \mathbb{R}^d}{\mathrm{argmin}} \sum_j P^\lambda_{i,j} \ C(H,H^t_j) $$
If the cost function $C$ corresponds to the squared $l_2$ distance, then the barycentric mapping reduces to the following form. 
$$\hat{H}_s = \text{diag}(P^\lambda 1_{n_t})^{-1} P^\lambda H_t$$
For domain adaptation, we typically choose the marginal distributions of the source and target node embeddings, $\mu_s$ and $\mu_t$, to be a  uniform distribution. This allows us to further simplify the equation as follows:
\begin{equation}
   \hat{H}_s = n_s P^\lambda H_t \label{eq_transform}
\end{equation}
where $n_s$ is the number of source nodes. 

Both $\hat{H}_s$ and $H_t$ are then fed to a fully connected layer to perform the node classification.
By replacing $H_s$ with $\hat{H}_s$ we train the model to work on data that has been transported to the target domain. This addresses the domain adaptation problem and allows the classifier to accurately classify the target nodes. 

Finally, the full OTGCN network is trained to minimize the following joint objective function:
\begin{eqnarray}
&& \mathcal{L} = \mathcal{L}_{CE} + \theta \mathcal{L}_{OT}\label{eq_loss_combined} 
\end{eqnarray}
where $\mathcal{L}_{CE}$ is the cross entropy loss given in Eqn. \eqref{eqn:crossent} and $\mathcal{L}_{OT}$ is the optimal transport loss defined as follows:
\begin{equation}
\mathcal{L}_{OT} = \langle P^{\lambda}, C\rangle - \lambda H(P^\lambda)
\end{equation}
with the hyperparameter $\theta$ controlling how much emphasis is placed on the optimal transport term. 


\section{Experimental Evaluation}

This section presents the experiments performed to evaluate the effectiveness of our proposed OTGCN framework. The source code for the framework and other aspects of our experiments can be found at \url{https://github.com/ajoystephens/otgcn}

\subsection{Data Preparation}

\begin{table}[b]
\caption{The data collection sites for ABIDE and its selection as source/target graph.}
\begin{center}
\begin{tabular}{|c|c|c|c|}
\hline
\textbf{Site} & \textbf{Samples} & \textbf{Autism} & \textbf{Dataset} \\ \hline
YALE & 41 & 22 & target \\ \hline
CALTECH & 15 & 5 & target \\ \hline
CMU & 11 & 6 & target \\ \hline
NYU & 172 & 74 & source \\ \hline
UM\_1 & 86 & 34 & source \\ \hline
USM & 67 & 43 & source \\ \hline
UCLA\_1 & 64 & 37 & source \\ \hline
PITT & 50 & 24 & source \\ \hline
MAX\_MUN & 46 & 19 & source \\ \hline
TRINITY & 44 & 19 & source \\ \hline
UM\_2 & 34 & 13 & source \\ \hline
KKI & 33 & 12 & source \\ \hline
LEUVEN\_2 & 28 & 12 & source \\ \hline
LEUVEN\_1 & 28 & 14 & source \\ \hline
OLIN & 28 & 14 & source \\ \hline
SDSU & 27 & 8 & source \\ \hline
SBL & 26 & 12 & source \\ \hline
STANFORD & 25 & 12 & source \\ \hline
OHSU & 25 & 12 & source \\ \hline
UCLA\_2 & 21 & 11 & source \\ \hline
\end{tabular}
\label{tab_abide_sites}
\end{center}
\end{table}

\begin{table}[t]
\caption{ABIDE Totals by Label and Dataset}
\begin{center}
\begin{tabular}{|c|c|c|c|}
\hline
 & \textbf{Autism} & \textbf{Control} & \textbf{Total} \\ \hline
\textbf{Source} & 370 (46.0\%) & 434 (54.0\%)& 804 \\ \hline
\textbf{Target} & 33 (49.3\%) & 34 (50.7\%) & 67 \\ \hline
\end{tabular}
\label{tab_abide_high_level}
\end{center}
\end{table}

The ABIDE \cite{dataset_abide} dataset is a combination of samples obtained from 20 different collection sites as shown in Table \ref{tab_abide_sites} for ASD classification. We split these collection sites into source and target sites for a total of 804 source samples and 67 target samples. Following the terminology used in other previous works in this area \cite{parisot2017spectral} \cite{mrida_wang2020multi} \cite{li2021braingnn} we refer to the two classes in the dataset as \emph{autism} and \emph{control}. Autism refers to a subject who developed ASD while control refers to a subject without ASD.

For the ABIDE dataset, the image data was prepared following the methodology of \cite{parisot2017spectral}. A population graph is constructed by considering each subject as a node and computing the similarity of their fMRI images for edge construction. Specifically, an edge was placed between a pair of nodes only if their image similarity exceeds certain threshold value. The edges are also weighted according to their computed similarity. The node features were derived from   phenotype information. After removing all columns with a known direct relationship to the class label, the features were one-hot-encoded as applicable. They were then normalized and any features with an abnormally high correlation to the label were also removed. 


\begin{figure*}[t]
\centerline{\includegraphics[width=\textwidth]{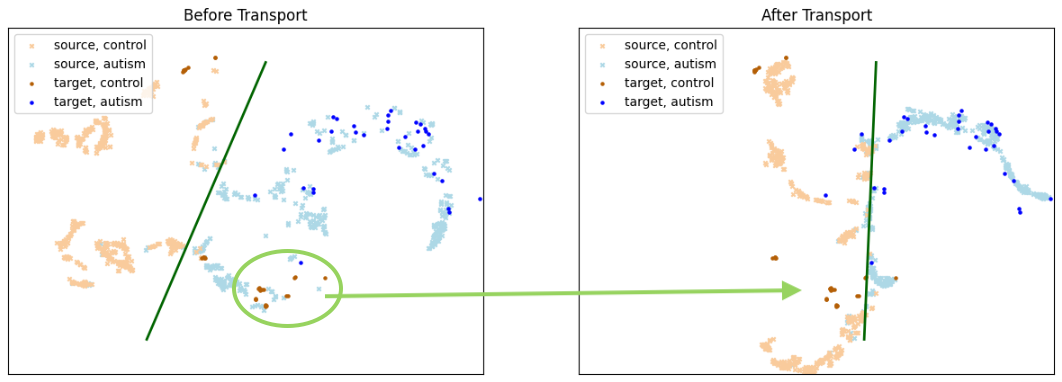}}
\caption{TSNE plot of combined embeddings before and after transport in the first pass of our OT layer. Points represent subjects and the lines are potential decision boundaries based off from each source dataset. The light colors represent the source dataset, before transport on the left and after transport on the right. The dark colors represent the target datset, which is the same in both cases. The green circle points out a portion of the target group which will likely be incorrectly classified before transportation occurs. The green arrow points to that same group on the right.}
\label{fig_transform_lines}
\end{figure*}

\subsection{Experimental Setup}

We compared the performance of OTGCN against the following baselines. Aside from GCN, the rest of the baseline methods are designed for cross network node classification problems.
\begin{itemize}
\item \textbf{GCN} \cite{gcn_kipf_welling}: A graph neural network that uses a message-passing technique to learn the feature embedding of the nodes. The architecture has been incorporated into OTGCN as well as other graph neural networks \cite{parisot2018disease} \cite{kazi2019inceptiongcn} \cite{jiang2020higcn}. The GCN implementation was written using code from the torch geometric library \cite{torch_geometric_2019} and involved two GCN layers followed by a fully connected classification layer.
\item \textbf{AdaGCN} \cite{cnnc_adagcn}: A CNNC technique which uses a series of GCN layers to learn separate embeddings for the source and target networks. It then performs domain adaptation by using an adversarial discriminator to force the two embedding into a shared domain. The AdaGCN implementation used in this paper was taken directly from the author's provide source code\footnote{\url{https://github.com/daiquanyu/AdaGCN_TKDE}}. No significant changes were made to the author's implementation, which consists of three GCN layers and a single layer discriminator. 
\item \textbf{ASN} \cite{cnnc_asn}: Another CNNC approach which consists of two separate GCN variational autoencoders (VAE) for the source and target, a shared GCN encoder which looks at both the source and target, and a adversarial discriminator. Here we used source code provide by the author\footnote{\url{https://github.com/yuntaodu/ASN}} with a small modification to correct for NaNs produced by the VAE reconstruction loss. The VAE reconstruction loss equation implemented in ASN includes a large exponential term followed by an aggregation step which can lead to NaNs if there is a large dataset or large values in VAE output. We addressed this issue by restricting values in VAE-generated embedding to between -10 and 10. Each of the three encoders in ASN were implemented with two GCN layers and the two decoders had a single layer. 
\item \textbf{ACDNE} \cite{cnnc_acdne}: A CNNC approach which uses special feature extraction layers to learn separate embeddings for the features and structure of each network. It then concatenates the two embeddings into a single source embedding and a single target embedding before sending both to an adversarial discriminator. We followed the authors source code\footnote{\url{https://github.com/shenxiaocam/ACDNE}} with this method as well and made no significant changes. Each of the four feature extractors consisted of two layers. 
\end{itemize}

For fair comparison we implemented OTGCN with two NFT layers and two GCN layers, similar to most of the baseline methods. We use the macro- and micro-F1 scores\cite{grandini2020metrics} as our evaluation metrics when comparing the performance of different classification methods.
We perform 10-fold cross validation to select the hyperparameters of OTGCN as well as the reported baselines. 
For OTGCN, we tune the hyperparameters for $\lambda$ and $\theta$ as well as standard hyperparameters such as learning rate. Possible hyperparameter values were [0.01, 0.03, 0.05] for $\lambda$ and [5, 10, 15] for $\theta$. Since the target graph is completely unlabeled and has potentially different distribution than the source graph, this poses a significant problem for hyperparameter tuning. 
To our knowledge there is no ideal solution to this problem. For this paper, the source dataset was split into 10 folds and the target dataset was \emph{excluded} from the hyperparameter tuning process. In each pass the fold selected for validation was removed from the remainder of the source dataset and treated as a distinct target dataset. We followed this process to choose the best hyperparameters for both our method and the reported baselines. 

For OTGCN the first and second NFT layers produced hidden embeddings of 32 and 64 units respectively. Similarly, the two GCN embeddings were also 32 and 64. The selected hyperparameters were $\lambda = 0.01$ and $\theta = 10$.

Each of ACDNE's feature extrators is build with two layer models which contain 64 and 32 hidden dimensions. We chose hyperparameter candidate values by  referencing the paper associated with the work. Our tuning script selected $1\times10^{-5}$ for the weight of the pairwise constraint loss and $1\times10^{-5}$ for the weight of the l2 regularization term. 

ADAGCN's three GCN layers have 64, 32 and 16 hidden dimensions and the discriminator has a single 16 unit hidden layer. We chose hyperparameter candidate values by looking at existing values within their source code for other datasets. There were three model-specific hyperparmeters for controlling various portion of their loss functions. Our tuning script selected $5\times10^{-5}$ for the weight of the l2 loss, 1.0 for the weight of the Wasserstein loss, and 5.0 for the weight of the penalty loss.

ASN's encoders were all two layers with 64 and 32 hidden units. Once again we chose hyperparameter candidate values by looking at existing information within the source code. Our tuning script chose 0.0001 for the weight of the difference loss, 1.0 for the weight of the domain loss, and 0.5 for the weight of the reconstruction loss.

After hyperparameter tuning each method was trained with the chosen hyperparameters ten times with distinct random seeds. Each time the model was trained on the entire labeled source dataset and evaluated on the target dataset. The mean and standard deviation of the resulting micro and macro F1 results are recorded in Table \ref{tab_results}.

\begin{table}[b]
\caption{Performance results on target dataset}
\begin{center}
\begin{tabular}{|c|c|c|}
\hline
\textbf{Method} & \textbf{Macro F1} & \textbf{Micro F1} \\ \hline
GCN & 0.50265 +/- 0.03876 & 0.55821 +/- 0.02234 \\ \hline
AdaGCN & 0.27853 +/- 0.18966 & 0.37112 +/- 0.22752 \\ \hline
ASN & 0.38712 +/- 0.09156 & 0.52537 +/- 0.04418 \\ \hline
ACDNE & 0.94310 +/- 0.04065 & 0.94328 +/- 0.04049 \\ \hline
OTGCN & \textbf{0.97907 +/- 0.00733} &\textbf{0.97910 +/- 0.00731} \\ \hline 

\end{tabular}
\label{tab_results}
\end{center}
\end{table}
\subsection{Results}



Table \ref{tab_results} shows the micro- and macro-F1 results for OTGCN and all baselines on the prepared ABIDE target dataset. These results demonstrate the improved performance of OTGCN over existing CNNC baselines. The next best performing result was from ACDNE, which did not rely on GCN for it's graph embedding, but rather extracted structural and feature information separately into separate embeddings. OTGCN and ACDNE perfomed significantly better than all other baselines, likely because these two methods do not rely strictly upon GCN and the homophily principle.  


Next we endeavor to establish the effectiveness of our optimal transport layer. To do this we save off a GCN embedding just before we send it to the very first OT transportation. We then transport the source embedding, use TSNE to reduce the embeddings to two dimensions and plot the results. 

Fig. \ref{fig_transform_lines} is an example of a plot after this process. Here the points are colored according to label and dataset. On the left we see source before transportation in lighter colors, autism is light blue and control is light orange. The target dataset is plotted with it in darker colors, autism is dark blue and control is dark orange. A dark green line illustrates a potential decision boundary according to the source dataset, and it is clear that it will likely misclassify a significant group of the target samples. On the right we see a similar plot, but in this case the source samples have been transported to the target domain via our OT layer. Once again a dark green line illustrates a potential decision boundary based off from the source samples. It is clear from this demonstration that a decision boundary derived from the transported source embedding will more accurately classify subjects in the target network.

\section{Conclusion}
This paper presents a deep learning framework called OTGCN to address the ASD detection problem using imaging and non-imaging information from multiple sites. OTGCN leverages ideas from graph neural networks to learn a feature representation of the nodes and optimal transport to effectively tackle the domain adaptation problem between source and target graphs. Our framework also incorporates a nonlinear feature transformation layer to alleviate the issue of graphs with low homophily. We experimentally compared the performance of OTGCN against several state of the art CNNC baselines using the multi-site ABIDE dataset. These experiments demonstrated the superior performance of OTGCN over the baseline methods. 

\section*{Acknowledgment}

This material is based upon work supported by the NSF
Program on Fairness in AI in collaboration with Amazon under
grant IIS-1939368. Any opinion, findings, and conclusions
or recommendations expressed in this material are those of
the author(s) and do not necessarily reflect the views of the
National Science Foundation or Amazon.

\bibliographystyle{IEEEtran}
\bibliography{refs}

\begin{thebibliography}{10}
\providecommand{\url}[1]{#1}
\csname url@samestyle\endcsname
\providecommand{\newblock}{\relax}
\providecommand{\bibinfo}[2]{#2}
\providecommand{\BIBentrySTDinterwordspacing}{\spaceskip=0pt\relax}
\providecommand{\BIBentryALTinterwordstretchfactor}{4}
\providecommand{\BIBentryALTinterwordspacing}{\spaceskip=\fontdimen2\font plus
\BIBentryALTinterwordstretchfactor\fontdimen3\font minus
  \fontdimen4\font\relax}
\providecommand{\BIBforeignlanguage}[2]{{%
\expandafter\ifx\csname l@#1\endcsname\relax
\typeout{** WARNING: IEEEtran.bst: No hyphenation pattern has been}%
\typeout{** loaded for the language `#1'. Using the pattern for}%
\typeout{** the default language instead.}%
\else
\language=\csname l@#1\endcsname
\fi
#2}}
\providecommand{\BIBdecl}{\relax}
\BIBdecl

\bibitem{lord2020autism}
C.~Lord, T.~S. Brugha, T.~Charman, J.~Cusack, G.~Dumas, T.~Frazier, E.~J.
  Jones, R.~M. Jones, A.~Pickles, M.~W. State \emph{et~al.}, ``Autism spectrum
  disorder,'' \emph{Nature reviews Disease primers}, vol.~6, no.~1, pp. 1--23,
  2020.

\bibitem{erkan2019autism}
U.~Erkan and D.~N. Thanh, ``Autism spectrum disorder detection with machine
  learning methods,'' \emph{Current Psychiatry Research and Reviews Formerly:
  Current Psychiatry Reviews}, vol.~15, no.~4, pp. 297--308, 2019.

\bibitem{zunino2018video}
A.~Zunino, P.~Morerio, A.~Cavallo, C.~Ansuini, J.~Podda, F.~Battaglia,
  E.~Veneselli, C.~Becchio, and V.~Murino, ``Video gesture analysis for autism
  spectrum disorder detection,'' in \emph{2018 24th international conference on
  pattern recognition (ICPR)}.\hskip 1em plus 0.5em minus 0.4em\relax IEEE,
  2018, pp. 3421--3426.

\bibitem{li2021braingnn}
X.~Li, Y.~Zhou, N.~Dvornek, M.~Zhang, S.~Gao, J.~Zhuang, D.~Scheinost, L.~H.
  Staib, P.~Ventola, and J.~S. Duncan, ``Braingnn: Interpretable brain graph
  neural network for fmri analysis,'' \emph{Medical Image Analysis}, vol.~74,
  p. 102233, 2021.

\bibitem{tong2017multi}
T.~Tong, K.~Gray, Q.~Gao, L.~Chen, D.~Rueckert, A.~D.~N. Initiative
  \emph{et~al.}, ``Multi-modal classification of alzheimer's disease using
  nonlinear graph fusion,'' \emph{Pattern recognition}, vol.~63, pp. 171--181,
  2017.

\bibitem{zeng2021multi}
Y.~Zeng, D.~Cao, X.~Wei, M.~Liu, Z.~Zhao, and Z.~Qin, ``Multi-modal relational
  graph for cross-modal video moment retrieval,'' in \emph{Proceedings of the
  IEEE/CVF Conference on Computer Vision and Pattern Recognition}, 2021, pp.
  2215--2224.

\bibitem{parisot2017spectral}
S.~Parisot, S.~I. Ktena, E.~Ferrante, M.~Lee, R.~G. Moreno, B.~Glocker, and
  D.~Rueckert, ``Spectral graph convolutions for population-based disease
  prediction,'' in \emph{Medical Image Computing and Computer Assisted
  Intervention- MICCAI 2017: 20th International Conference, Quebec City, QC,
  Canada, September 11-13, 2017, Proceedings, Part III 20}.\hskip 1em plus
  0.5em minus 0.4em\relax Springer, 2017, pp. 177--185.

\bibitem{parisot2018disease}
S.~Parisot, S.~I. Ktena, E.~Ferrante, M.~Lee, R.~Guerrero, B.~Glocker, and
  D.~Rueckert, ``Disease prediction using graph convolutional networks:
  application to autism spectrum disorder and alzheimer’s disease,''
  \emph{Medical image analysis}, vol.~48, pp. 117--130, 2018.

\bibitem{kazi2019inceptiongcn}
A.~Kazi, S.~Shekarforoush, S.~Arvind~Krishna, H.~Burwinkel, G.~Vivar,
  K.~Kort{\"u}m, S.-A. Ahmadi, S.~Albarqouni, and N.~Navab, ``Inceptiongcn:
  receptive field aware graph convolutional network for disease prediction,''
  in \emph{Information Processing in Medical Imaging: 26th International
  Conference, IPMI 2019, Hong Kong, China, June 2--7, 2019, Proceedings
  26}.\hskip 1em plus 0.5em minus 0.4em\relax Springer, 2019, pp. 73--85.

\bibitem{jiang2020higcn}
H.~Jiang, P.~Cao, M.~Xu, J.~Yang, and O.~Zaiane, ``Hi-gcn: A hierarchical graph
  convolution network for graph embedding learning of brain network and brain
  disorders prediction,'' \emph{Computers in Biology and Medicine}, vol. 127,
  p. 104096, 2020.

\bibitem{tan2018survey}
C.~Tan, F.~Sun, T.~Kong, W.~Zhang, C.~Yang, and C.~Liu, ``A survey on deep
  transfer learning,'' in \emph{Artificial Neural Networks and Machine
  Learning--ICANN 2018: 27th International Conference on Artificial Neural
  Networks}.\hskip 1em plus 0.5em minus 0.4em\relax Springer, 2018, pp.
  270--279.

\bibitem{dataset_abide}
C.~Craddock, S.~Sikka, B.~Cheung, R.~Khanuja, S.~S. Ghosh, C.~Yan, Q.~Li,
  D.~Lurie, J.~Vogelstein, R.~Burns \emph{et~al.}, ``Towards automated analysis
  of connectomes: The configurable pipeline for the analysis of connectomes
  (c-pac),'' \emph{Front Neuroinform}, vol.~42, pp. 10--3389, 2013.

\bibitem{gcn_kipf_welling}
T.~N. Kipf and M.~Welling, ``Semi-supervised classification with graph
  convolutional networks,'' \emph{arXiv preprint arXiv:1609.02907}, 2016.

\bibitem{mcpherson2001birds}
M.~McPherson, L.~Smith-Lovin, and J.~M. Cook, ``Birds of a feather: Homophily
  in social networks,'' \emph{Annual review of sociology}, vol.~27, no.~1, pp.
  415--444, 2001.

\bibitem{yuan2016autism}
J.~Yuan, C.~Holtz, T.~Smith, and J.~Luo, ``Autism spectrum disorder detection
  from semi-structured and unstructured medical data,'' \emph{EURASIP Journal
  on Bioinformatics and Systems Biology}, vol. 2017, pp. 1--9, 2016.

\bibitem{carette2018automatic}
R.~Carette, F.~Cilia, G.~Dequen, J.~Bosche, J.-L. Guerin, and L.~Vandromme,
  ``Automatic autism spectrum disorder detection thanks to eye-tracking and
  neural network-based approach,'' in \emph{Internet of Things (IoT)
  Technologies for HealthCare: 4th International Conference, HealthyIoT 2017,
  Angers, France, October 24-25, 2017, Proceedings 4}.\hskip 1em plus 0.5em
  minus 0.4em\relax Springer, 2018, pp. 75--81.

\bibitem{nogay2020machine}
H.~S. Nogay and H.~Adeli, ``Machine learning (ml) for the diagnosis of autism
  spectrum disorder (asd) using brain imaging,'' \emph{Reviews in the
  Neurosciences}, vol.~31, no.~8, pp. 825--841, 2020.

\bibitem{mrida_cheng2017multi}
B.~Cheng, M.~Liu, D.~Shen, Z.~Li, D.~Zhang, and A.~D.~N. Initiative.,
  ``Multi-domain transfer learning for early diagnosis of alzheimer’s
  disease,'' \emph{Neuroinformatics}, vol.~15, pp. 115--132, 2017.

\bibitem{mrida_wang2020multi}
J.~Wang, L.~Zhang, Q.~Wang, L.~Chen, J.~Shi, X.~Chen, Z.~Li, and D.~Shen,
  ``Multi-class asd classification based on functional connectivity and
  functional correlation tensor via multi-source domain adaptation and
  multi-view sparse representation,'' \emph{IEEE transactions on medical
  imaging}, vol.~39, no.~10, pp. 3137--3147, 2020.

\bibitem{mrida_guan2021multi}
H.~Guan, L.~Wang, and M.~Liu, ``Multi-source domain adaptation via optimal
  transport for brain dementia identification,'' in \emph{2021 IEEE 18th
  International Symposium on Biomedical Imaging (ISBI)}.\hskip 1em plus 0.5em
  minus 0.4em\relax IEEE, 2021, pp. 1514--1517.

\bibitem{zhang2021survey}
C.~Zhang, Y.~Xie, H.~Bai, B.~Yu, W.~Li, and Y.~Gao, ``A survey on federated
  learning,'' \emph{Knowledge-Based Systems}, vol. 216, p. 106775, 2021.

\bibitem{li2020multi}
X.~Li, Y.~Gu, N.~Dvornek, L.~H. Staib, P.~Ventola, and J.~S. Duncan,
  ``Multi-site fmri analysis using privacy-preserving federated learning and
  domain adaptation: Abide results,'' \emph{Medical Image Analysis}, vol.~65,
  p. 101765, 2020.

\bibitem{wang2022metateacher}
Z.~Wang, M.~Ye, X.~Zhu, L.~Peng, L.~Tian, and Y.~Zhu, ``Metateacher:
  Coordinating multi-model domain adaptation for medical image
  classification,'' \emph{Advances in Neural Information Processing Systems},
  vol.~35, pp. 20\,823--20\,837, 2022.

\bibitem{cnnc_cdne}
X.~Shen, Q.~Dai, S.~Mao, F.-l. Chung, and K.-S. Choi, ``Network together: Node
  classification via cross-network deep network embedding,'' \emph{IEEE
  Transactions on Neural Networks and Learning Systems}, vol.~32, no.~5, pp.
  1935--1948, 2020.

\bibitem{cnnc_acdne}
X.~Shen, Q.~Dai, F.-l. Chung, W.~Lu, and K.-S. Choi, ``Adversarial deep network
  embedding for cross-network node classification,'' in \emph{Proceedings of
  the AAAI Conference on Artificial Intelligence}, vol.~34, no.~03, 2020, pp.
  2991--2999.

\bibitem{cnnc_adagcn}
Q.~Dai, X.~Shen, X.-M. Wu, and D.~Wang, ``Network transfer learning via
  adversarial domain adaptation with graph convolution,'' \emph{arXiv preprint
  arXiv:1909.01541}, 2019.

\bibitem{cnnc_udagcn}
M.~Wu, S.~Pan, C.~Zhou, X.~Chang, and X.~Zhu, ``Unsupervised domain adaptive
  graph convolutional networks,'' in \emph{Proceedings of The Web Conference
  2020}, 2020, pp. 1457--1467.

\bibitem{cnnc_asn}
X.~Zhang, Y.~Du, R.~Xie, and C.~Wang, ``Adversarial separation network for
  cross-network node classification,'' in \emph{Proceedings of the 30th ACM
  International Conference on Information \& Knowledge Management}, 2021, pp.
  2618--2626.

\bibitem{monge1781memoire}
G.~Monge, ``M{\'e}moire sur la th{\'e}orie des d{\'e}blais et des remblais,''
  \emph{Mem. Math. Phys. Acad. Royale Sci.}, pp. 666--704, 1781.

\bibitem{kantorovich2006translocation}
L.~V. Kantorovich, ``On the translocation of masses,'' \emph{Journal of
  mathematical sciences}, vol. 133, no.~4, pp. 1381--1382, 2006.

\bibitem{ot_kantorovitch}
\BIBentryALTinterwordspacing
L.~Kantorovitch, ``On the translocation of masses,'' \emph{Management Science},
  vol.~5, no.~1, pp. 1--4, 1958. [Online]. Available:
  \url{https://doi.org/10.1287/mnsc.5.1.1}
\BIBentrySTDinterwordspacing

\bibitem{sinkhorn_cuturi}
\BIBentryALTinterwordspacing
M.~Cuturi, ``Sinkhorn distances: Lightspeed computation of optimal transport,''
  in \emph{Advances in Neural Information Processing Systems}, C.~Burges,
  L.~Bottou, M.~Welling, Z.~Ghahramani, and K.~Weinberger, Eds., vol.~26.\hskip
  1em plus 0.5em minus 0.4em\relax Curran Associates, Inc., 2013. [Online].
  Available:
  \url{https://proceedings.neurips.cc/paper/2013/file/af21d0c97db2e27e13572cbf59eb343d-Paper.pdf}
\BIBentrySTDinterwordspacing

\bibitem{ot_map__courty}
N.~Courty, R.~Flamary, and D.~Tuia, ``Domain adaptation with regularized
  optimal transport,'' in \emph{Machine Learning and Knowledge Discovery in
  Databases: European Conference, ECML PKDD 2014, Nancy, France, September
  15-19, 2014. Proceedings, Part I 14}.\hskip 1em plus 0.5em minus 0.4em\relax
  Springer, 2014, pp. 274--289.

\bibitem{ot_map__ferradans}
S.~Ferradans, N.~Papadakis, G.~Peyr{\'e}, and J.-F. Aujol, ``Regularized
  discrete optimal transport,'' \emph{SIAM Journal on Imaging Sciences},
  vol.~7, no.~3, pp. 1853--1882, 2014.

\bibitem{torch_geometric_2019}
M.~Fey and J.~E. Lenssen, ``Fast graph representation learning with {PyTorch
  Geometric},'' in \emph{ICLR Workshop on Representation Learning on Graphs and
  Manifolds}, 2019.

\bibitem{grandini2020metrics}
M.~Grandini, E.~Bagli, and G.~Visani, ``Metrics for multi-class classification:
  an overview,'' \emph{arXiv preprint arXiv:2008.05756}, 2020.

\end{thebibliography}

\end{document}